\documentclass{article}
\usepackage[T1]{fontenc}

\makeatletter

\newcommand{\LyX}{L\kern-.1667em\lower.25em\hbox{Y}\kern-.125emX\@}

\makeatother

\begin{document}

{\par\centering \textbf{\LARGE Generalised quantum anharmonic oscillator using
an operator ordering approach} \par}
\vspace{2cm}

{\par\centering {\large Anirban Pathak }\large \par}

{\par\centering {\large Department of Physics, }\large \par}

{\par\centering {\large Visva-Bharati, }\large \par}

{\par\centering {\large Santiniketan-731235, }\large \par}

{\par\centering {\large India}\large \par}

\begin{abstract}
We construct a generalised expression for the normal ordering of \( (a+a^{\dagger })^{m} \)
for integral values of \( m \) and use the result to study the quantum anharmonic
oscillator problem in the Heisenberg approach. In particular, we derive generalised
expressions for energy eigen values and frequency shifts for the Hamiltonian
\( H=\frac{x^{2}}{2}+\frac{\dot{x}^{2}}{2}+\frac{\lambda }{m}x^{m} \). We also
derive a closed form first order multi scale perturbation theoretic operator
solution of this Hamiltonian with a view to generalise some recent results of
Bender and Bettencourt .
\\

\end{abstract}

\section{Introduction}

The simple harmonic oscillator (SHO) model can be used to understand a wide
variety of physical phenomena ranging from problems in Newtonian mechanics to
those in quantum field theory. This solvable model describes small oscillation
of a system about the mean position of equilibrium under the action of a linear
restoring force. However, for a real physical problem one has to incorporate
the anharmonicity in the model Hamiltonian. In this case the equation of motion
is not exactly solvable and, in-fact, the anharmonic oscillators (AHO) provide
one of the simplest examples of quantum mechanical systems which can not be
solved without making use of approximation techniques. This was why AHO problem
got considerable attention of the physicists and mathematicians from the very
beginning of the subject. It is an outstanding curiosity to note that studies
in the AHO problems have enriched the subject of large order perturbation theory
{[}\ref{bendern1}-\ref{lowdin}{]}, divergent expansion of quantum mechanics
{[}\ref{simonn2}-\ref{graffi}{]}, Laplace transformation representation of
energy eigen values {[}\ref{ivanov}-\ref{fernandez}{]}, computational physics,
Pade and Borel summation of perturbation series {[}\ref{borel-pade}{]}. 

In the present paper we shall study the generalised quantum anharmonic oscillator
problem in the Heisenberg representation by making use of two simple normal
ordering theorems derived by us for the expansion of \( (a+a^{\dagger })^{m} \).
We shall see that the merit of the present paper is its simplicity. For example,
the present method provides, on the one hand, a straight forward mathematical
frame work to construct expressions for the energy eigen values as well as frequency
shifts and, on the other hand, generalises some of the recent results on the
topic by Bender and Bettencourt {[}\ref{Bender1}-\ref{Bender2}{]}. 

The Hamiltonian of a generalised anharmonic oscillator having unit mass and
unit frequency is written as 
\begin{equation}
\label{ten}
\begin{array}{lcl}
H & = & H_{0}+\frac{\lambda }{m}x^{m}\\
 & = & \frac{x^{2}}{2}+\frac{\dot{x}^{2}}{2}+\frac{\lambda }{m}x^{m}
\end{array}
\end{equation}
 with
\begin{equation}
\label{11}
x=\frac{1}{\sqrt{2}}(a^{\dagger }+a)\, \, .
\end{equation}
Here we have chosen to work in units in which \( \hbar =1. \) Equation of motion
corresponding to (\ref{ten}) is 
\begin{equation}
\label{eqm}
\ddot{x}+x+\lambda x^{m-1}=0.
\end{equation}
 Equation (\ref{eqm}) can not be solved exactly for \( m>2 \). However, large
number of approximation methods are available for solving (\ref{eqm}) for particular
values of \( m. \) These include perturbation technique {[}\ref{Nayfeh}{]},
variation of parameters and Taylor series approach {[}\ref{Margenau}{]}. Ordinary
perturbation technique leads to the unwanted secular terms that grows up rapidly
with time. Secular terms are unwanted because they are in conflict with the
physical requirement that the solution be finite. There are some methods which
are successfully used to sum up the secular terms for all orders. Multiscale
perturbation theory (MSPT) is one of those methods {[}\ref{Bender1}-\ref{Nayfeh}{]}.

Anharmonic oscillator problem can be approached in two different ways in quantum
mechanics. One way is the c-number approach or the Schroedinger approach and
the other way is the operator approach or the Heisenberg approach. There is
an extensive amount of literature on c-number approach {[}\ref{bendern1}-\ref{borel-pade},\ref{C-num}{]}
but the situation is different for the Heisenberg approach because the complicated
operator algebra tends to pose serious mathematical problems. Our main objective
in this work is to provide a theoretical frame work which is free from calculational
difficulties. The first operator solution of a quantum oscillator was given
by Bender and Bettencourt {[}\ref{Bender1},\ref{Bender2}{]} using MSPT. They
generalised the existing theory of MSPT into an operator approach to get the
zeroth order solution involving a quantum operator analog of the classical first
order frequency shift and interpreted it as an operator mass renormalization
that expresses the first order shift of all energy levels. Somewhat later Mandal
{[}\ref{Mandal}{]} approached the problem (up to first order) using Taylor
series method. All these solutions were found for the quartic oscillator only.
Application of the Taylor series method to sextic and octic oscillator problems
are now in order {[}\ref{Pathak}{]}. 

The Schroedinger approach to AHO problem in quantum mechanics has made a significant
contribution for the development of the perturbative techniques as we use today
but the suitability of the other approach (Heisenberg approach) has not yet
been studied in detail. Till now we don't have any operator solutions of higher
anharmonic oscillators although these might be relevant to various physical
problems. Keeping these in mind, we will suggest a novel method in which MSPT
results for anharmonic oscillators in general can be found. To that end we prove
in section 2 two theorems to construct a normal ordered expansion of \( (a+a^{\dagger })^{m} \)
. We obtain a generalised expression for the energy eigen values in section
3. We devote section 4 to construct a generalised solution for the equation
of motion (\ref{eqm}) and specialise our result to reproduce some of the existing
results as useful checks on the generalised solution obtained by us.

\section{Operator Ordering theorems}

On a very general ground one knows that in quantum mechanics and quantum field
theory proper ordering of the operators plays a crucial role. Let \( f(a,a^{\dagger }) \)
be an arbitrary operator function of the usual bosonic annihilation and creation
operators \( a \) and \( a^{\dagger } \) which satisfy the commutation relation
\begin{equation}
\label{one}
[a,a^{\dagger }]=1.
\end{equation}
One can write \( f(a,a^{\dagger }) \) in such a way that all powers of \( a^{\dagger } \)
always appear to the left of all powers of \( a \). Then \( f(a,a^{\dagger }) \)
is said to be normal ordered. In this paper we want to write \( (a+a^{\dagger })^{m} \)
in the normal ordered form for integral values of \( m \). Traditionally for
a given value of \( m \) this is achieved by using a very lengthy procedure
which involves repeated application of (\ref{one}). One of our objectives in
this work is to construct a normal ordered expansion of \( (a+a^{\dagger })^{m} \)
without taking recourse to such repeated applications. 

We shall denote normal ordered form of \( f \) by \( f_{N} \). On the other
hand \( :\, f\, : \) will denote an operator obtained from \( f \) by arranging
all powers of \( a^{\dagger } \) to the left of all powers of \( a \) without
making use of the commutation relation in (\ref{one}). Now if \( f=aa^{\dagger } \)
then \( f_{N}=a^{\dagger }a+1 \) and \( :\, f\, :=a^{\dagger }a \). Therefore,
we can write
\begin{equation}
\label{two}
:\, (a+a^{\dagger })^{m}:=:\, (a^{\dagger }+a)^{m}:=a^{m}+^{m}C_{1}a^{\dagger }a^{m-1}+...+^{m}C_{r}a^{\dagger ^{r}}a^{m-r}+..+a^{\dagger ^{m}}.
\end{equation}
 Thus in this notation \( :\, (a+a^{\dagger })^{m}: \) is simply a binomial
expansion in which powers of the \( a^{\dagger } \) are always kept to the
left of the powers of the \( a \). To write \( (a^{\dagger }+a)_{N}^{m} \)
we shall proceed by using the following theorems.

\subsection{Theorem 1. }

\begin{equation}
\label{th1.1}
:\, (a^{\dagger }+a)^{m}:\, (a^{\dagger }+a)=:\, (a^{\dagger }+a)^{m+1}:+m\, :\, (a^{\dagger }+a)^{m-1}:\, \, .
\end{equation}

Proof: From (\ref{two}) \( :\, (a^{\dagger }+a)^{m}:\, (a^{\dagger }+a) \)
can be written in the form 
\begin{equation}
\label{four}
\begin{array}{cl}
 & :\, (a^{\dagger }+a)^{m}:\, (a^{\dagger }+a)\\
= & \left[ a^{\dagger ^{m+1}}+(\, ^{m}C_{1}+^{m}C_{0})a^{\dagger ^{m}}a+(\, ^{m}C_{2}+^{m}C_{1})a^{\dagger ^{m-1}}a^{2}+....+(\, ^{m}C_{r}+^{m}C_{r-1})a^{\dagger ^{m-r+1}}a^{r}+...\right] \\
+ & (\, ^{m}C_{1}a^{\dagger ^{m-1}}+2\, ^{m}C_{2}a^{\dagger ^{m-2}}a+...+r\, ^{m}C_{r}a^{\dagger ^{m-r}}a^{r-1}+...)\, .\\

\end{array}
\end{equation}

The theorem in (\ref{th1.1}) can be obtained by simply using the following
identities in the above 
\begin{equation}
\label{five}
\begin{array}{lcl}
\, \, \, \, \, \, (a^{r}a^{\dagger })_{N} & = & a^{\dagger }a^{r}+ra^{r-1},\\
\, \, \, \, \, \, r\, ^{n}C_{r} & = & n\, ^{n-1}C_{r-1},\\
and\, \, \, \, ^{m+1}C_{r+1} & = & (\, ^{m}C_{r}+^{m}C_{r+1})\, .
\end{array}
\end{equation}
 Note that first identity in (\ref{five}) can be proved with the help of the
general operator ordering theorems {[}\ref{Louisell}{]} while the other two
are trivial.

\subsection{Theorem 2 : }

For any integral values of \( m \) 
\begin{equation}
\label{th2.1}
(a^{\dagger }+a)_{N}^{m}=\sum ^{m}_{r=0,2,4..}t_{r}\, ^{m}C_{r}:\, (a^{\dagger }+a)^{m-r}:
\end{equation}
with
\begin{equation}
\label{th2.2}
t_{r}=\frac{(r-1)!}{2^{(\frac{r}{2}-1)}(\frac{r}{2}-1)!}\, \, for\, r\geq 4
\end{equation}
 and 
\begin{equation}
\label{th2.3}
t_{0}=t_{2}=1\, .
\end{equation}

Proof:

Using theorem 1, we can write
\begin{equation}
\label{six}
\begin{array}{lcl}
(a^{\dagger }+a)_{N} & = & :\, (a^{\dagger }+a):\\
(a^{\dagger }+a)_{N}^{2} & = & :\, (a^{\dagger }+a)^{2}:+1\\
(a^{\dagger }+a)_{N}^{3} & = & :\, (a^{\dagger }+a)^{3}:+3:\, (a^{\dagger }+a):\\
(a^{\dagger }+a)_{N}^{4} & = & :\, (a^{\dagger }+a)^{4}:+6\, :\, (a^{\dagger }+a)^{2}:+3\\
(a^{\dagger }+a)_{N}^{5} & = & :\, (a^{\dagger }+a)^{5}:+10\, :\, (a^{\dagger }+a)^{3}:+15\, :\, (a^{\dagger }+a):\\
\, \, \, \, \, \, \, \, ... & ... & \, \, \, \, \, \, \, \, \, \, \, \, \, \, \, \, \, \, \, \, \, \, \, \, \, \, \, \, ...\\
\, \, \, \, \, \, \, \, ... & ... & \, \, \, \, \, \, \, \, \, \, \, \, \, \, \, \, \, \, \, \, \, \, \, \, \, \, \, \, ...\\
\, \, \, \, \, \, \, \, ... & ... & \, \, \, \, \, \, \, \, \, \, \, \, \, \, \, \, \, \, \, \, \, \, \, \, \, \, \, \, ...\\
(a^{\dagger }+a)_{N}^{9} & = & :\, (a^{\dagger }+a)^{9}:+36:\, (a^{\dagger }+a)^{7}:+378:\, (a^{\dagger }+a)^{5}:\\
 & + & 1260:\, (a^{\dagger }+a)^{3}:+945:\, (a^{\dagger }+a):\, \, .\\
 &  & \\
 &  & 
\end{array}
\end{equation}
 From (\ref{six}) we venture to identify the general form of the above expansion
for a given value of \( m \) as
\begin{equation}
\label{seven}
(a^{\dagger }+a)_{N}^{m}=\sum ^{m}_{r=0,2,4..}t_{r}\, ^{m}C_{r}:\, (a^{\dagger }+a)^{m-r}:\, \, .
\end{equation}
It is easy to check that (\ref{seven}) gives all the expansions of (\ref{six})
such that (\ref{seven}) is true for \( m=1,2,..,9 \). The general validity
of (\ref{seven}) can be ensured by using the method of induction. This establishes
that (\ref{seven}) gives the normal order expansion of \( (a^{\dagger }+a)^{m} \)
for any arbitrary integer \( m \). We shall now use this normal ordered expansion
to study the anharmonic oscillator problem.

\section{Energy eigen values }

The first order energy eigen value \( E_{1} \) is \( <n|H|n> \), where \( |n> \)
is the number state. In the literature there are two lengthy procedures {[}\ref{Powell}{]}
to obtain \( E_{1} \). The first one is the usual normal ordering method. This
method involves iterative use of (\ref{one}) and number of iteration increases
very first as \( m \) increases. For a given value of \( m \) we need   \( \left[ 2^{m}-(m+1)\right]  \)
iterations. This shows that for large \( m \) the construction of the expansion
 \( (a^{\dagger }+a)_{N}^{m} \) becomes formidable. In the second procedure
one proceeds by making repeated application of \( x \) operator on the number
state. As in the normal ordering method this procedure is also equally lengthy.
In the following we implement theorem 2 to derive an uncomplicated method to
construct an expansion for \( E_{1} \) for the Hamiltonian (\ref{ten}).

Using the result in (\ref{th2.1}) we can write the expression for \( E_{1} \)
for any integer value of \( m \) in the form
\begin{equation}
\label{12}
\begin{array}{lcl}
E_{1} & = & (n+\frac{1}{2})+\frac{\lambda }{2^{\frac{m}{2}}m}<n|\sum ^{m}_{r=0,2,4..}t_{r}\, ^{m}C_{r}:\, (a^{\dagger }+a)^{m-r}:\, |n>\\
 & = & (n+\frac{1}{2})+\frac{\lambda }{2^{\frac{m}{2}}m}<n|\sum ^{m}_{r=0,2,4..}t_{r}\, ^{m}C_{r}\, ^{m-r}C_{\frac{m-r}{2}}\, a^{\dagger ^{\frac{m-r}{2}}}a^{\frac{m-r}{2}}|n>\\
 & = & (n+\frac{1}{2})+\frac{\lambda }{2^{\frac{m}{2}}m}\sum ^{m}_{r=0,2,4..}t_{r}\, ^{m}C_{r}\, ^{m-r}C_{\frac{m-r}{2}}\, ^{n}C_{\frac{m-r}{2}}\, (\frac{m-r}{2})!
\end{array}
\end{equation}
 The expression (\ref{12}) involves summations which are easy to evaluate and
thereby avoids the difficulties associated with earlier iterative procedures.
Although somewhat forced, it may be tempting to compare the simplicity sought
in our approach with the use of logarithm in a numerical calculation or use
of integral transforms in solving a partial differential equation. It is of
interest to note that expression similar to (\ref{12}) can also be constructed
for the first order energy eigen value of a Hamiltonian in which the anharmonic
term is a polynomial in \( x \). This type of Hamiltonians are very frequently
used in nonlinear optics.

\section{MSPT solution of the generalised quantum anharmonic oscillator.}

We want to obtain MSPT operator solution of equation (\ref{eqm}) for arbitrary
integral values of \( m \). The essential idea behind our approach is that
the quantum operator analog of the classical first order frequency shift is
an operator function ( \( \Omega (H_{0}) \) ) of the unperturbed Hamiltonian
\( H_{0} \) and secondly a correct solution should reproduce the first order
energy spectrum. Now the general form of the zeroth order solution of (\ref{eqm})
is given by 
\begin{equation}
\label{sol1}
\begin{array}{lcl}
x_{0}(t) & = & \frac{1}{G(n)}\left[ x(0)cos(t+\lambda \Omega (H_{0})t)+cos(t+\lambda \Omega (H_{0})t)x(0)\right. \\
 & + & \left. \dot{x}(0)sin(t+\lambda \Omega (H_{0})t)+sin(t+\lambda \Omega (H_{0})t)\dot{x}(0)\right] 
\end{array}
\end{equation}
 where \( G(n) \) is a normalization factor. So our task is to find out \( \Omega (H_{0}) \)
and \( G(n) \) in general. From equation (\ref{12}) we obtain the energy difference
for two consecutive energy levels as 
\begin{equation}
\label{sol2}
\begin{array}{lcl}
\omega _{n,n-1} & = & (E_{1})_{m,n}-(E_{1})_{m,n-1}\\
 & = & 1+\lambda \omega (m,n)\\
 & = & 1+\frac{\lambda }{2^{\frac{m}{2}}m}\sum ^{(m-2)}_{r=0,2,4..}t_{r}\, ^{m}C_{r}\, ^{m-r}C_{\frac{m-r}{2}}\, ^{n-1}C_{\frac{m-r-2}{2}}\, (\frac{m-r}{2})!\, .
\end{array}
\end{equation}

Since the correct quantum operator solution has to give the first order energy
spectrum, we should have
\begin{equation}
\label{sol3}
<n-1|x_{0}(t)|n>=<n-1|x_{0}(0)|n>cos[t+\lambda \omega (m,n)t]+<n-1|\, p_{0}(0)|n>sin[t+\lambda \omega (m,n)t]\, .
\end{equation}
 Equation (\ref{sol1}) and (\ref{sol3}) impose restrictions on our unknown
functions \( \Omega (H_{0}) \) and \( G(n) \). The condition imposed on \( \Omega (H_{0}) \)
is

\begin{equation}
\label{sol4}
<n|\Omega (H_{0})|n>+<n-1|\Omega (H_{0})|n-1>=2\omega (m,n)
\end{equation}
or, 
\begin{equation}
\label{sol5}
\Omega (n+\frac{1}{2})+\Omega (n-\frac{1}{2})=2\omega (m,n)\, \, .
\end{equation}
For a particular \( m \) right hand side is a known polynomial in \( n \)
and our job is simply to find out \( \Omega (n+\frac{1}{2}) \). We obtain this
as 
\begin{equation}
\label{sol6}
\Omega (n+\frac{1}{2})=2\left[ \sum ^{n}_{k=0}(-1)^{n-k}\omega (m,k)\right] +(-1)^{n+\frac{m}{2}}\frac{t_{m}}{2^{\frac{m-2}{2}}m}.
\end{equation}
 Substituting the functional form of \( \Omega (n+\frac{1}{2}) \) or \( \Omega (H_{0}) \)
in (\ref{sol1}) if we impose condition (\ref{sol3}) we will get 
\begin{equation}
\label{sol6.1}
G(n)=2cos\left[ \frac{\lambda t}{2}\left( \Omega (n+\frac{1}{2})-\Omega (n-\frac{1}{2})\right) \right] .
\end{equation}

The results in (\ref{sol6}) and (\ref{sol6.1}) when substituted in (\ref{sol1})
solve the generalised quantum anharmonic oscillator problem.

\subsection{Specific results and their comparison with the existing spectra:}

Although we have solved the quantum anharmonic oscillator in general, it is
not possible to compare our solution directly with other existing results since
the present study happens to be the first operator solution of the generalised
anharmonic oscillator. There are methods to calculate first order classical
frequency shift for particular \( m \) in the appropriate classical limit (\( x(0)=a,\, \dot{x}(0)=0 \)).
Here we calculate some specific results from our general expressions and compare
them with the existing results.

For \( m=4 \) we have, 
\begin{equation}
\label{quartic1}
\Omega (n+\frac{1}{2})=\frac{3n}{4}+\frac{3}{8}=\frac{3}{4}(n+\frac{1}{2})\, .
\end{equation}
 Therefore, 
\begin{equation}
\label{sol8}
\begin{array}{lcl}
\, \, \, \, \, \, \Omega (H_{0}) & = & \frac{3}{4}H_{0}\\
and\, \, \, \, \, G(n) & = & 2cos(\frac{3\lambda t}{8})\, .
\end{array}
\end{equation}
 In terms of (\ref{sol8}) the total solution for the quantum quartic anharmonic
oscillator is 
\begin{equation}
\label{quartic2}
\begin{array}{lcl}
x(t)|_{m=4} & = & \frac{1}{2cos(\frac{3\lambda t}{8})}\left[ x(0)cos[t+\frac{3\lambda t}{4}H_{0}]+cos[t+\frac{3\lambda t}{4}H_{0}]x(0)\right. \\
 & + & \left. \dot{x}(0)sin[t+\frac{3\lambda t}{4}H_{0}]+sin[t+\frac{3\lambda t}{4}H_{0}]\dot{x}(0)\right] 
\end{array}
\end{equation}
 This exactly coincides with the solutions given in {[}\ref{Bender1},\ref{Bender2}
and \ref{Mandal}{]} and gives the correct classical frequency shift {[}\ref{bdr}{]}
in the limit \( x(0)=a,\, \dot{x}(0)=0 \). Similarly we have 
\begin{equation}
\label{sol9}
\begin{array}{lcl}
x(t)|_{m=6} & = & \frac{1}{2cos(\frac{5\lambda t}{4}n)}\left[ x(0)cos[t+\frac{5\lambda }{4}(H_{0}^{2}+\frac{1}{4})t]+cos[t+\frac{5\lambda }{4}(H_{0}^{2}+\frac{1}{4})t]x(0)\right. \\
 & + & \left. \dot{x}(0)sin[t+\frac{5\lambda }{4}(H_{0}^{2}+\frac{1}{4})t]+sin[t+\frac{5\lambda }{4}(H_{0}^{2}+\frac{1}{4})t]\dot{x}(0)\right] 
\end{array},
\end{equation}
 
\begin{equation}
\label{sol10}
\begin{array}{lcl}
x(t)|_{m=8} & = & \frac{1}{2cos[\frac{35\lambda }{64}(6n^{2}+3)t]}\left[ x(0)cos[t+\frac{35\lambda }{64}(4H_{0}^{3}+5H_{0})t]+cos[t+\frac{35\lambda }{64}(4H_{0}^{3}+5H_{0})t]x(0)\right. \\
 & + & \left. \dot{x}(0)sin[t+\frac{35\lambda }{64}(4H_{0}^{3}+5H_{0})t]+sin[t+\frac{35\lambda }{64}(4H_{0}^{3}+5H_{0})t]\dot{x}(0)\right] 
\end{array},
\end{equation}
and 
\begin{equation}
\label{sol11}
\begin{array}{lcl}
x(t)|_{m=10} & = & \frac{1}{2cos[\frac{63\lambda t}{8}(n^{3}+2n)]}\left[ x(0)cos[t+\frac{63\lambda }{16}(H_{0}^{4}+\frac{7}{2}H^{2}_{0}+\frac{9}{16})t]\right. \\
 & + & cos[t+\frac{63\lambda }{16}(H_{0}^{4}+\frac{7}{2}H^{2}_{0}+\frac{9}{16})t]x(0)+\dot{x}(0)sin[t+\frac{63\lambda }{16}(H_{0}^{4}+\frac{7}{2}H^{2}_{0}+\frac{9}{16})t]\\
 & + & \left. sin[t+\frac{35\lambda }{64}\frac{63\lambda }{16}(H_{0}^{4}+\frac{7}{2}H^{2}_{0}+\frac{9}{16})t]\dot{x}(0)\right] \, .
\end{array}
\end{equation}
 The solutions for sextic and octic oscillator exactly coincides with the solution
obtained by us using Taylor series approach {[}\ref{Pathak}{]} and all the
solutions give the correct classical frequency shifts in the appropriate limit
{[} \ref{R. Dutt},\ref{Bradbury}{]}.

\section{Summary and concluding remarks}

We conclude by noting that depending on the nature of nonlinearity in a physical
problem the treatment of higher anharmonic oscillators assumes significance.
But studies in such oscillators (for \( m>4 \)) are not undertaken in the Heisenberg
approach presumably because the existing methods tends to introduce inordinate
mathematical complications in a detailed study. In the present work we contemplate
to circumvent them by proving a theorem for the expansion of \( (a^{\dagger }+a)_{N}^{m} \).
Thus the results of the present work are expected to serve a useful purpose
for physicists working in nonlinear mechanics, molecular physics, quantum optics
and quantum field theory. 
\vspace{1.5cm}

\textbf{Acknowledgments}: \emph{The author is thankful to the CSIR for the award
of a Junior Research Fellowship. He is also grateful to Prof B K Talukdar and
Dr S Mandal for their kind interest in this work.~}\\
\\
\\

\textbf{References}

\begin{enumerate}
\item \label{bendern1} Bender C M and Wu T T, \emph{Phys Rev}., \textbf{184} (1969)
1231
\item \label{bendern2}Bender C M and Wu T T, \emph{Phys Rev. Let} , \textbf{27} (1971)
461
\item \label{bendern3}Bender C M and Wu T T, \emph{Phys Rev}. \emph{D}, \textbf{7}(1973)
1620
\item \label{Areta}Arteca G A, Fernandez F M and Castro E A, (1990) Large Order Perturbation
Theory and Summation Methods in Quantum Mechanics (Berlin: Springer-Verlag).
\item \label{lowdin}Lowdin P O (Ed.) \emph{Int. J. Quant. Chem}(1982) ,( Proceedings
of the Sanibel Workshop on Perturbation Theory at large order, Sanibel Conference,
Florida, 1981)
\item \label{simonn2}Simon B, \emph{Int. J. Quant. Chem.} \textbf{21} (1982) 3
\item \label{graffi}Graffi S, Greechi V and Simon B, \emph{Phys Lett.} \textbf{32B}
(1970) 631.
\item \label{ivanov}Ivanov I A, \emph{J. Phys A} \textbf{31} (1998) 5697; Ivanov
I A, \emph{J. Phys A} \textbf{31} (1998) 6995.
\item \label{weniger}Weniger E J, \emph{Annals of Physics} \textbf{246} (1996) 133.
\item \label{weniger2}Weniger E J, Cizek J and Vinette F, \emph{J. Math. Phys}. \textbf{39}
(1993) 571.
\item \label{vinette}Vienette F and Cizek J, \emph{J. Math. Phys.} \textbf{32} (1991)
3392
\item \label{fernandez}Fernandez F and Cizek J, \emph{Phys. Lett.} \textbf{166A}
(1992) 173.
\item \label{borel-pade}Graffi S, Greechi V and Turchetti G, \emph{IL Nuovo Cimento}
\textbf{4B} (1971) 313
\item \label{Bender1}Bender C M, Bettencourt L M A, \emph{Phys. Rev. Lett.} \textbf{77}
(1996) 4114.
\item \label{Bender2}Bender C M, Bettencourt L M A, \emph{Phys. Rev. D} 54 (1996)
1710.
\item \label{Nayfeh}Nayfeh A H (1981) \emph{Introduction to Perturbation Techniques}
(Wiely, New York).
\item \label{Margenau}Margenau H and Murphy G M (1956) \emph{The Mathematics of Physics
and Chemistry} (D. Van Nostrand Company, Princeton).
\item \label{C-num}Bender C M, \emph{J. Math. Phys.} \textbf{11} (1970) 796; Biswas
S N etal \emph{J. Math. Phys}. \textbf{14} (1973) 1190; Killingbeck J, \emph{Rep.
Prog. Phys.} \textbf{40} (1977) 963; ; Zamatsila J, Cizek J and Skala L \emph{Annals
of Physics} \textbf{276} (1998) 39.
\item \label{Mandal} Mandal S (1998), \emph{J.Phys. A} \textbf{31} L501.
\item \label{Pathak} Pathak A and Mandal S, Communicated for publication.
\item \label{Louisell}Louisell W H (1973) Quantum Statistical Properties of Radiation
( John Wiley and Sons, New York).
\item \label{Powell}Powel J L and Crasemann B (1961) \emph{Quantum Mechanics} (Addision-Wesley,
Massachusetts).
\item \label{bdr} Bhaumik K and Dutta-Ray B (1975) \emph{J. Math. Phys}. \textbf{16}
1131.
\item \label{R. Dutt}Dutt R, Lakshmanan M, \emph{J.Math.Phys}. \textbf{17} (1976)
482.
\item \label{Bradbury}Bradbury T C and Brintzenhoff A (1971) \emph{J. Math. Phys}.
\textbf{12} 1269.
\end{enumerate}
\end{document}